\documentclass[aps,pra,reprint, amsmath, amssymb,superscriptaddress]{revtex4-1}

\usepackage{bm}
\usepackage[retainorgcmds]{IEEEtrantools}
\usepackage{graphicx}
\usepackage{mathrsfs}
\usepackage{amsmath}
\usepackage{amssymb}
\usepackage{color}
\usepackage{amsfonts}
\usepackage{nicefrac}

\newcommand{\ud}{\,\mathrm{d}}

\DeclareMathOperator{\sech}{sech}
\DeclareMathOperator{\tr}{tr}

\newcommand{\floor}[1]{\left\lfloor #1 \right\rfloor}

\begin{document}

\title{Energy barriers between metastable states in first-order quantum phase transitions}
\date{\today}
\author{Sascha Wald}
\email{swald@sissa.it}
\affiliation{SISSA - International School for Advanced Studies, via Bonomea 265, I-34136 Trieste}
\affiliation{Groupe de Physique Statistique, D\'epartement de Physique de la Mati\`ere et des Mat\'eriaux, Institut Jean Lamour (CNRS UMR 7198), Universit\'e de Lorraine Nancy, B.P. 70239,
F - 54506 Vand\oe uvre l\`es Nancy Cedex, France}
\affiliation{Theoretische Physik, Universit{\"a}t des Saarlandes, D-66123 Saarbr\"ucken, Germany}
\author{Andr\'e M. Timpanaro}
\affiliation{Centro de Matem\'atica, Computa\c c\~ao e Cogni\c c\~ao da Universidade Federal do ABC,  09210-580 Santo Andr\'e, Brazil}
\author{Cecilia Cormick}
\affiliation{Instituto de F\'isica Enrique Gaviola, CONICET and Universidad Nacional de C\'ordoba, Ciudad Universitaria, X5016LAE, C\'ordoba, Argentina}
\author{Gabriel T. Landi}
\email{gtlandi@if.usp.br}
\affiliation{Instituto de F\'isica da Universidade de S\~ao Paulo,  05314-970 S\~ao Paulo, Brazil}

\begin{abstract}
A system of neutral atoms trapped in an optical lattice and dispersively coupled to the field of an optical cavity can 
realize a variation of the Bose-Hubbard model with infinite-range interactions. This model exhibits a first order quantum 
phase transition between a Mott insulator and a charge density wave, with spontaneous symmetry breaking between even and 
odd sites, as was recently observed experimentally [Landig \emph{et. al.}, Nature {\bf 532} (2016)]. 
In the present paper we approach the analysis of this transition using a variational model which allows us to establish the notion 
of an energy barrier separating the two phases. Using a discrete WKB method we then show that the local tunneling of atoms 
between adjacent sites lowers this energy barrier and hence facilitates the transition.  Within our simplified description, 
we are thus able to augment the phase diagram of the model with information concerning the height of the barrier separating 
the metastable minima from the global minimum in each phase, which is an essential aspect for the
understanding of the reconfiguration dynamics induced by a quench across a quantum 
critical point.

\end{abstract}
\maketitle{}

%
%
\section{\label{sec:int}Introduction}
%
%

Quantum phase transitions are driven by a competition between different terms in the Hamiltonian, which lead to a rearrangement of the many-body ground-state  as a certain parameter is varied. 
Despite being known for many decades, this subject has seen a boom of interest in recent years, chiefly due to  experimental progress in condensed matter physics and ultra-cold atoms in optical lattices \cite{Greiner2002,Baumann2010,Islam2015,Leonard2017a,Leonard2017}. 
Theoretical and computational advances have also helped improve our understanding of the basic mechanisms underlying these transitions \cite{Sachdev1998}.
In addition to understanding the equilibrium properties of quantum phase transitions, considerable effort is currently put into understanding properties related to the relaxation towards equilibrium, for instance after a quantum quench \cite{Calabrese2016,Mitra2017,Wald2015b,Wald2017}. 
This requires one to understand how the system reconfigures itself from one phase to another as the critical point is traversed. 
In classical phase transitions, this reconfiguration  is generated by  thermal fluctuations, which induce a coarsening dynamics of nucleated domains \cite{HenkelBook1,Henkel2010,Maraga2015,Chandran2013,Sciolla2013,Chiocchetta2017}.
In the case of quantum phase transitions, on the other hand, one obviously expects that it will be the quantum fluctuations 
that assume this role. 
Notwithstanding, there are still several
open questions about how exactly this takes place.

First order (discontinuous) transitions are of special interest due to their inherent hysteretic behavior, which means that the system must overcome an extensive energy barrier in order to traverse  from one phase to another. 
A clear demonstration of this behavior was recently given in Refs.~\cite{Landig2016,Hruby2017} for a model of interacting bosons in an optical lattice.
The model consists of a variation of the standard two-dimensional Bose-Hubbard model \cite{Fisher1989,Jaksch1998,Sachdev1998,Greiner2002} 
that includes a long-range interaction between atoms in the different checkerboard sub-lattices of a square lattice. 
It presents in total four quantum phases:
 superfluid (SF),  supersolid (SS),  Mott insulator (MI) and  charge density wave (CDW). 
We shall concentrate, in particular, on the MI-CDW transition, which is of first order. 
Detailed theoretical investigations of this model were recently given in a variety of papers \cite{Chen2016a,Liao2017,Flottat2017b,Sundar2016,Dogra2016b,Niederle2016,Panas2016}. 
However, these papers were mostly concerned with the equilibrium phase diagram. 
To our knowledge, the mechanisms which allow the transition between 
the MI and CDW phases and the description of the barrier separating these phases, have so far not been explored, 
despite being  crucial for further studies on the relaxation (quench) dynamics.
The purpose of this paper is to shed  light on these mechanisms, which should serve as a first step towards more complex
non-equilibrium studies.



The MI and CDW phases \cite{Hruby2017} are illustrated in the insets in Fig.~\ref{fig:landau}. 
In the former, the atoms tend to distribute uniformly throughout the square lattice, whereas in the latter they tend
to aggregate into either the even or the odd sub-lattice.
%
Assuming thermal fluctuations can be neglected, the only available mechanism that allows the system to reconfigure from one phase to the other is the local tunneling of atoms between adjacent sites. 
The tunneling should therefore affect the energy landscape of the system, lowering the barrier separating the two phases. 
In this paper we discuss how this effect occurs. 
We begin by considering the situation with no tunneling, in which case the phases can be described exactly by a Landau theory for the imbalance order parameter.
Next we introduce the effects of tunneling using a  variational Ansatz based on a set of representative states between the MI and CDW phases. 
As we show, within this variational setting we can model the effect of the hopping as a general tight-binding Hamiltonian on the space of even-odd imbalances. Then, using a discrete WKB method, we are able to study how the local tunneling affects the barrier between metastable states. 
Our main result is an equilibrium phase diagram augmented with information on the height of the energy barrier, shown in Fig.~\ref{fig:PD}.
The equilibrium phases we obtain match well with other analyses in the literature \cite{Flottat2017b,Dogra2016b} and the information on the energy barrier heights agree qualitatively with  recent experimental results \cite{Hruby2017}.  

%
%
\section{\label{sec:landau}Landau theory}
%
%

We consider here a bosonic system on a 2D square lattice with $K$ sites, each site described by a bosonic operator $b_i$. 
We write the Hamiltonian as \cite{Landig2016}
\begin{equation}\label{H}
H = H_0 + \mathcal{T},
\end{equation}
where 
\begin{equation}\label{H0}
H_0 = \sum\limits_{i=1}^K \frac{U_s}{2} \hat{n}_i (\hat{n}_i - 1) - \frac{U_\ell}{K} \Theta^2,
\end{equation}
and 
\begin{equation}\label{T}
\mathcal{T} = - J \sum\limits_{\langle i,j\rangle} ( b_i^\dagger b_j + b_j^\dagger b_i).
\end{equation}
In these expressions, $U_s$ is the short-ranged repulsive interaction, $J$ is the hopping parameter, $\hat{n}_i = b_i^\dagger b_i$ and $U_\ell$ is the long-range interaction.
The sum in Eq.~(\ref{T}) is over all nearest neighbors in the square lattice.
Moreover, we divide the lattice into a checkerboard structure consisting of even and odd sites, so that the nearest neighbors of an even site are always odd and vice-versa. 
Then, finally, the operator $\Theta$ appearing in Eq~(\ref{H0}) is defined as 
\begin{equation}\label{Theta}
\Theta = \sum_{i\in e} \hat{n}_i - \sum_{i\in o} \hat{n}_i,
\end{equation}  
which we refer to as  the imbalance operator between the even ($e$) and odd ($o$) sub-lattices.
In this paper we work entirely in the canonical ensemble, with a fixed number of particles $N$. 
For simplicity, we also consider a spatially homogeneous system.

The long-range term $U_\ell$ introduces a long-ranged checkerboard interaction which favors an imbalanced occupation of the even and odd sub-lattices. 
When $U_\ell = 0$  we recover the usual Bose-Hubbard model \cite{Fisher1989,Jaksch1998,Sachdev1998,Greiner2002}, which presents a quantum phase transition between a  SF and a MI.
The presence of the checkerboard interaction introduces, in addition to these phases, a SS 
and a CDW phase.
The MI and CDW phases appear 
when  $J/U_s \ll 1$ and are thus essentially determined by the competition between $U_s$ and $U_\ell$. 
For  $J\to0$ the Hamiltonian becomes diagonal in the Fock basis and the partition function may be computed exactly at zero temperature, in both the canonical and the grand canonical ensembles
(ensemble inequivalence in long-range interacting systems forces us to explicitly specify the ensemble we work in \cite{Campa2009}).

The partition function in the canonical ensemble is
\begin{equation}
Z = \sum\limits_{\text{states(N)}} e^{-\beta H_0},
\end{equation}
where the sum is over all states $(n_1,\ldots,n_K)$ with the constraint that $\sum_i n_i = N$ .
Note that, in this case all operators in $H_0$ commute, so we do not need to differentiate between operators and their eigenvalues. 
We may write $Z$ as
\begin{equation}
Z = \sum\limits_{\Theta=-N}^N \Bigg\{ e^{\beta U_\ell \Theta^2/K} \sum\limits_{\text{states}(N,\Theta)} e^{-\beta \frac{ U_s}{2} \sum\limits_i n_i(n_i-1)}\Bigg\}.
\end{equation}
Next we note that fixing $N$ and $\Theta$ is tantamount to fixing the occupations $N_e$ and $N_o$ of the even and odd sub-lattices; viz., 
\begin{equation}
N_e = \frac{N+\Theta}{2}, \qquad\qquad N_o = \frac{N-\Theta}{2}.
\end{equation}
Thus we may write $Z$ as 
\begin{equation}
Z=\sum\limits_{\Theta=-N}^N  e^{\beta U_\ell \Theta^2/K}  \zeta(N_e) \zeta(N_o),
\end{equation}
where
\begin{equation}
\zeta(N_x) = \sum\limits_{\text{states}_x(N_x)} e^{-\frac{\beta U_s}{2} \sum\limits_{i\in x} n_i(n_i-1)}.
\end{equation}
Here the notation $\text{states}_x(N_x)$ means a sum only over sub-lattice $x = e,o$ with a fixed number of particles $N_x$. 

Next we define $\theta = \Theta/K$, $\rho = N/K$ and $\rho_x = 2 N_x/K$ and also
the quantity $\phi(\rho_x)$ according to $\zeta(N_x) = e^{-(\beta K/2) \phi(\rho_x)}$. 
Then, in the thermodynamic limit,  the partition function may be written as 
\begin{equation}
Z =\text{const} \times \int\limits_{-\rho}^\rho \ud \theta  \; e^{- \beta K f(\theta)},
\end{equation}
with the Landau free energy 
\begin{equation}\label{landau_gen}
f(\theta) = - U_\ell \theta^2  + \frac{\phi(\rho+\theta) + \phi(\rho-\theta)}{2},
\end{equation}
where we used the fact that $\rho_e = \rho + \theta$ and $\rho_o = \rho-\theta$. 
Thus, we conclude that in the limit of zero hopping ($J=0$), the system is described \emph{exactly} by a Landau theory, both at zero and at finite temperatures. 
This is of course not surprising given the mean-field character of the long-range interaction. 

The quantities $\phi(\rho_x)$ may be determined exactly in the limit of zero temperature
\begin{equation}\label{phi_min}
\phi(\rho_x) = \frac{U_s}{K} \min_{N_x} \bigg\{ \sum\limits_i n_i(n_i-1)\bigg\}.
\end{equation}
The minimization is equivalent to a Bose-Hubbard model with no hopping, so the configuration which minimizes this quantity will be the one
which is as close as possible to the Mott Insulator.
Let 
\begin{equation}\label{xi}
N'_x = N_x - \frac{K}{2}\floor{\rho_x},
\end{equation}
where $\floor{x}$ is the floor function. 
Then there will be $N_x'$ sites with occupation $\floor{\rho_x}+1$ and the remainder with occupation $\floor{\rho_x}$. 
Thus we get 
\begin{equation}\label{phi}
\phi(\rho_x) = U_s \floor{\rho_x} \bigg\{ \rho_x - \frac{1}{2} (1 + \floor{\rho_x})\bigg\}.
\end{equation}
Substituting this in Eq.~(\ref{landau_gen}) then yields the free energy as a function of the filling $\rho$ and the imbalance $\theta$. 

\begin{figure}
\centering
\includegraphics[width=0.4\textwidth]{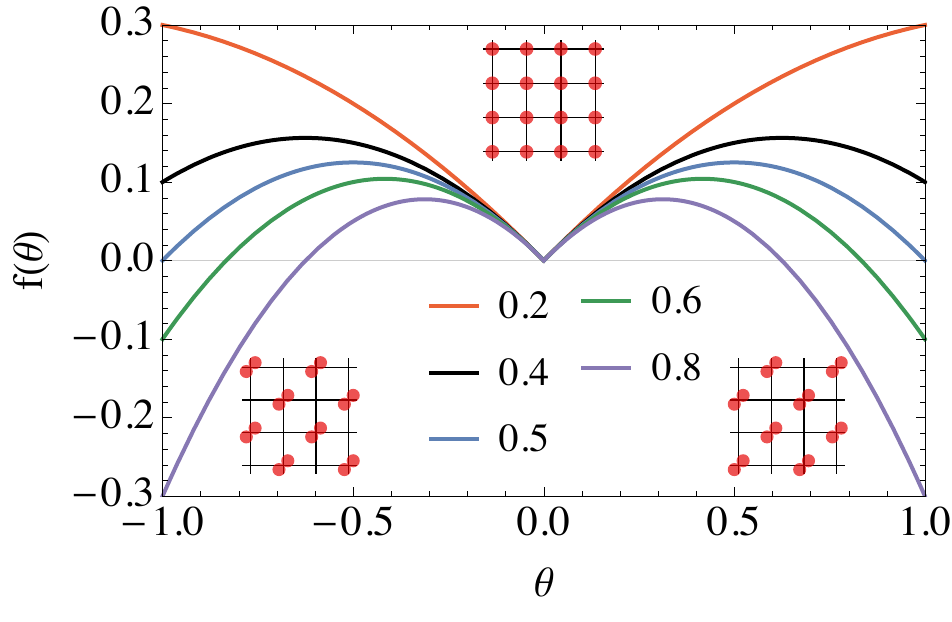}
\caption{\label{fig:landau} 
(Color Online) The Landau free energy~(\ref{landau}), Eq.~(\ref{landau}),  for filling $\rho = N/K = 1$ and different values of $U_\ell/U_s$, ranging from 0.2 (uppermost curve) to 0.8 (lowermost).
The system undergoes a first order quantum phase transition at $U_\ell/U_s = 1/2$, from a Mott Insulator (MI) corresponding to a minimum at $\theta = 0$  to a Charge Density Wave (CDW) state whose minimum is at $\theta = \pm 1$. 
}
\end{figure}

Eq.~(\ref{landau_gen}) simplifies considerably when $\rho = 1$. 
In this case, for $\theta \in [-1,1]$ it  follows that 
\[
\phi(1+\theta) + \phi(1-\theta) = U_s |\theta|.
\]
Thus we get 
\begin{equation}\label{landau}
f(\theta) = - U_\ell \theta^2 + \frac{U_s}{2}|\theta|,
\end{equation}
which is a remarkably simple expression. 
This result is shown in Fig.~\ref{fig:landau} for different values of $U_\ell/U_s$. 
As can be seen, the system presents three minima at $\theta = 0$ (MI) and $\theta = \pm1$ (CDW).
The global minimum changes at $U_\ell = U_s/2$, signaling a first order quantum phase transition.

\begin{figure}
\centering
\includegraphics[width=0.4\textwidth]{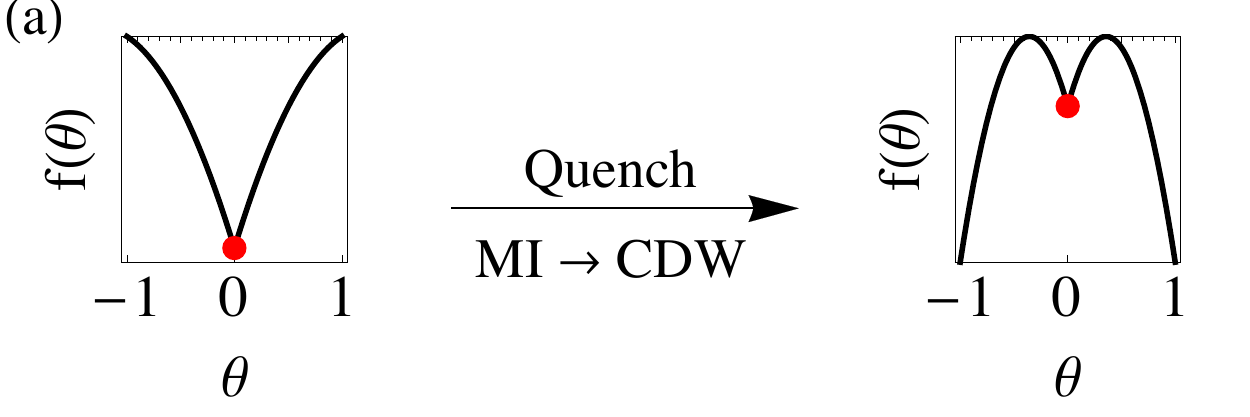}\\
\includegraphics[width=0.4\textwidth]{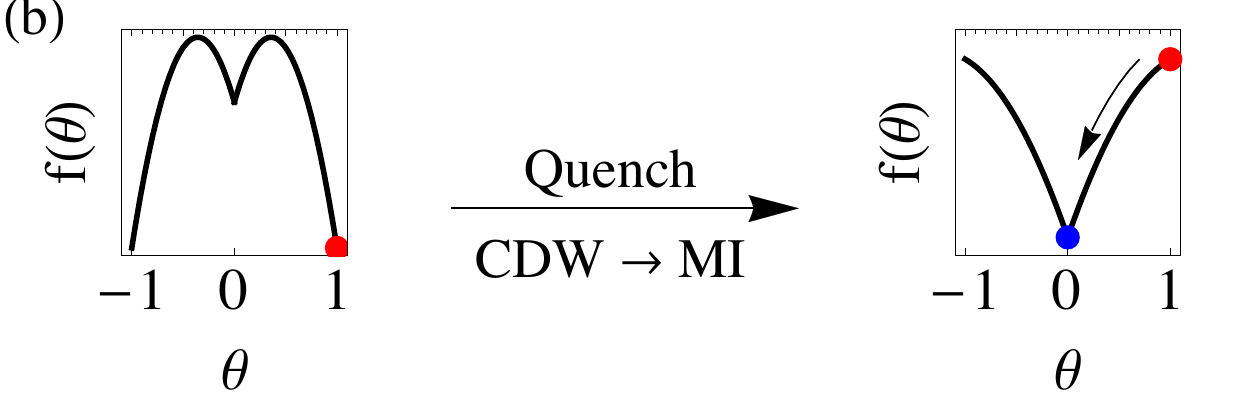}
\caption{\label{fig:hysteresis} 
(Color Online) Description of the asymmetric behavior in the hysteresis of the MI-CDW transition. 
Image (a) describes the  free energy as a
function of $\theta$ when the system was initially prepared in the MI state ( $U_\ell/U_s = 0.2$) and then suddenly quenched to the CDW phase ($U_\ell/U_s = 0.7$). 
In this case, there is a stronger resistance for the system to reconfigure towards the CDW minimum since the MI state continues to be a local minimum. 
Conversely, on image (b) we show the reverse effect. Now, after the quench, the system is found in a configuration which is no longer a local minima and, therefore, can easily reconfigure itself to the MI phase. 
}
\end{figure}

An interesting point, which has important consequences for the phase reconfiguration, is that the MI minimum at $\theta = 0$ always exists, whereas the CDW minima at $\theta = \pm 1$ cease to exist when $U_\ell < U_s/4$, as illustrated in Fig.~\ref{fig:landau}. 
This asymmetry can manifest in physical observables e.g. during a quench or a hysteresis loop protocol, as was indeed observed experimentally in \cite{Landig2016}.
The basic idea is explained diagrammatically  in Fig.~\ref{fig:hysteresis}. 
Suppose the system is initially prepared in the MI state and is then suddenly quenched to the CDW phase. 
In this case, the MI state will continue to be a local minimum. As long as there is a small energy barrier to surmount, there will be a natural resistance for the system to reconfigure 
(this is analogous to the Stoner and Wolfarth model in magnetism \cite{Stoner1948}). 
Conversely, if we prepare the system in the CDW phase and then quench to the MI, as in Fig.~\ref{fig:hysteresis}(b), then the CDW state will no longer be a local minimum and there will be no energetic cost for the system to reconfigure to the MI phase. 
As a consequence, if one measures the hysteresis loop as $U_\ell$ is varied between the two phases, the result will be inherently asymmetric. 
Of course, this analysis does not take into account thermal or quantum fluctuations, nor the important fact that a quench implies injecting excitations into the system. 
Notwithstanding, it shows the existence of an inherent asymmetry between the two phases.

\begin{figure}
\centering
\includegraphics[width=0.22\textwidth]{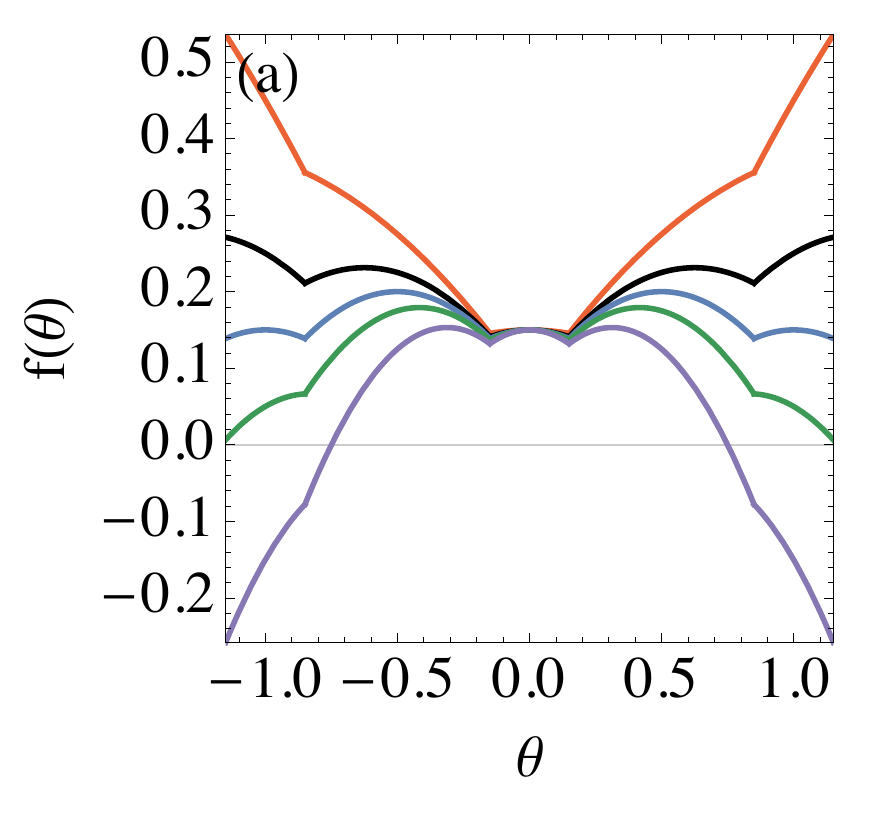}\quad
\includegraphics[width=0.22\textwidth]{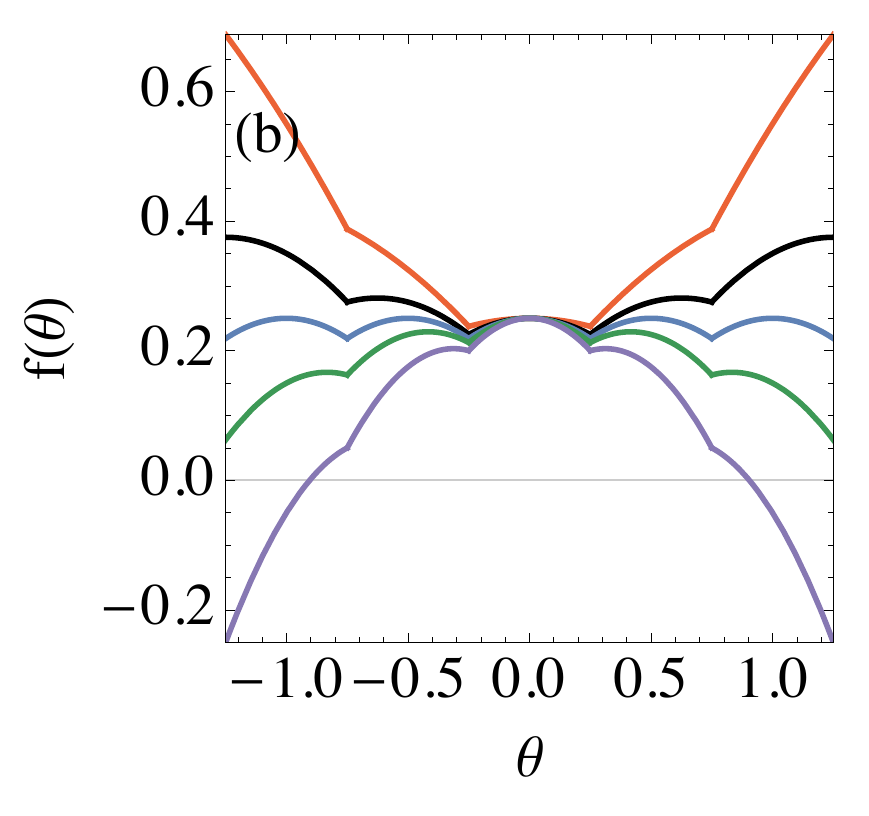}\\
\includegraphics[width=0.22\textwidth]{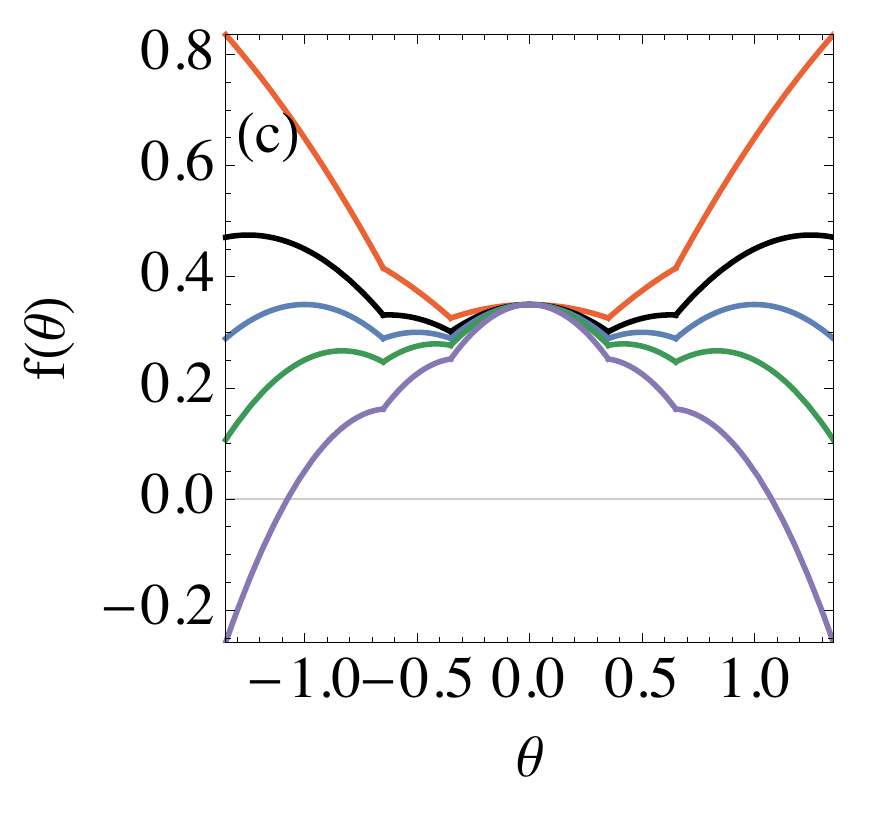}\quad
\includegraphics[width=0.22\textwidth]{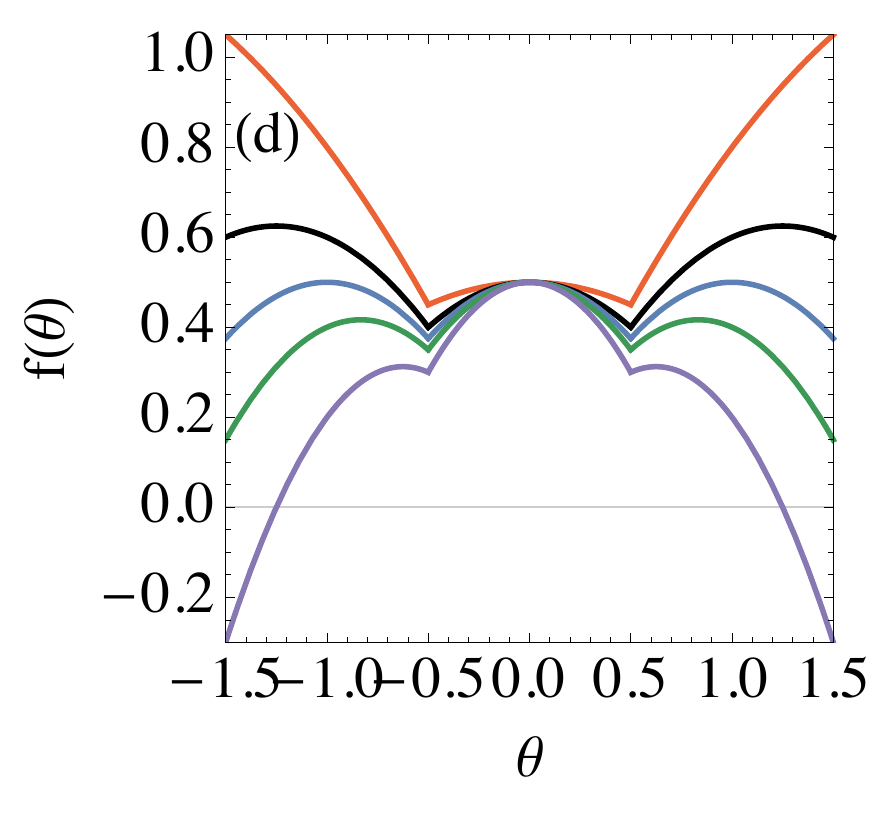}
\caption{\label{fig:landau_filling}
The Landau free energy~(\ref{landau_gen}) for different values of $U_\ell$.
Each image corresponds to a different filling $\rho$:
(a) 1.15, (b) 1.25, (c) 1.35 and (d) 1.5.
The values of $U_\ell$ and the color scheme are the same as in Fig.~\ref{fig:landau}.
}
\end{figure}

The Landau free energy presented in Fig.~\ref{fig:landau} corresponds to the case $\rho = 1$ [cf. Eq.~(\ref{landau})] and shows one or three minima. 
For non-integer fillings other local minima appear.  
Examples of Eq.~(\ref{landau_gen}) for different fillings $\rho$ are  shown in Fig.~\ref{fig:landau_filling}. 
As can be seen, non-integer fillings produce local minima at intermediate values of $\theta$. 
The existence of these degenerate configurations can be traced back to the results in Eq.~(\ref{xi}) for the occupations that minimize the energy. 
We conjecture that these intermediate configurations might explain some of the plateaus observed in Fig.~4 of Ref.~\cite{Hruby2017}

%
%
\section{\label{sec:var}Variational treatment of the hopping term}
%
%

When $J \neq 0$ the Hamiltonian~(\ref{H}) can no longer be diagonalized exactly. 
Instead, we approach the problem  using a variational Schr\"odinger Lagrangian Ansatz
\begin{equation}\label{Lagrangian}
\mathcal{L} = \langle \psi |( i \partial_t - H) | \psi\rangle,
\end{equation}
with a  choice of variational wave-function $|\psi\rangle$ that conveniently captures the physics of the MI-CDW transition. For concreteness,  we fix $\rho = 1$ so that the most relevant states are of the form $|\text{MI}\rangle = |1, \ldots, 1; 1, \ldots, 1 \rangle$,  $|\text{CDW}_e\rangle=|2, \ldots, 2;0, \ldots, 0\rangle$ and $|\text{CDW}_o\rangle=|0, \ldots, 0;2, \ldots, 2\rangle$, where the semi-colon separates the Fock occupations of the even and odd sub-lattices. 

Our goal is to describe the idea of an energy barrier forming between the MI and CDW phases. 
In order to do so, we need to choose a representative set of states which describe relevant intermediate configurations. Since we are interested in the case of small tunneling, we discard states with more than double occupation, restricting for each site to the occupations $n_i = 0,1,2$. 
States with higher occupation will have a much higher $U_s$-repulsion, with no gain in $U_\ell$-attraction, and hence should be energetically suppressed as long as $J$ is small.

This truncation of the total Hilbert space is still insufficient to render the problem tractable, and we still need to find an appropriate choice of suitable intermediate states between MI and CDW. A 
natural choice are the \emph{states of definite imbalance}; that is, eigenstates of the imbalance operator $\Theta$ in Eq.~(\ref{Theta}).
Since $\rho=1$, the eigenvalues $Q$ of $\Theta$ can vary from $-K$ to $K$ in steps of two, with MI being $Q=0$ and CDW being $Q = \pm K$. 
Of course, each eigenvalue is highly degenerate and may thus be labeled as $|Q,\nu\rangle$, for some additional index $\nu$. 
We then have 
\begin{equation}
\Theta |Q,\nu\rangle = Q |Q,\nu\rangle.
\end{equation}
The MI and CDW states are the only elements of the $Q$ states which are not degenerate. 
In particular, $|\text{MI}\rangle = |Q = 0\rangle$ and $|\text{CDW}_{e,o} \rangle = |Q = \pm K\rangle$. 

In order to make the problem tractable, we must choose a representative set of states for intermediate values of $Q$, which interpolate smoothly between these configurations. These states shall be denoted simply by $|Q\rangle$. 
The principles that will guide our specific choice will be discussed below. 
With this choice, the Sch\"odinger Lagrangian will then produce a reduced dynamical description within this truncated subspace. 
The main results of the remainder of this section will be the matrix elements of the Hamiltonian~(\ref{H}) in this $Q$ basis.
The Hamiltonian $H_0$ in~(\ref{H0}) will be diagonal, while the hopping term 
will only connect neighboring $|Q\rangle$ states. 
With these matrix elements at hand, we will then construct the effective 
Hamiltonian 
\begin{equation}\label{Heff}
\langle \psi | H | \psi \rangle = \sum\limits_Q  \bigg\{ \epsilon_Q \psi_Q^* \psi_Q + \gamma_Q^+ \psi_{Q+2}^* \psi_Q + \gamma_Q^- \psi_{Q-2}^* \psi_Q \bigg\},
\end{equation}
with coefficients $\epsilon_Q, \gamma_Q^\pm$ to be found in the following and 
with
\begin{equation}
|\psi\rangle = \sum\limits_Q \psi_Q |Q\rangle. 
\end{equation}
Hence,  within this variational picture the Hamiltonian is tridiagonal and the 
hopping term produces an effective tight-binding model within the space of 
imbalances $Q$. 

We now proceed to derive these results in detail. 
In Sec. \ref{sec:ana} they will be used to study the resulting properties of 
the system. 
 
\subsection{The choice of $Q$ states}

Recall that we are focusing on $\rho = N/K = 1$, and restricting to occupations $n_i = 0,1,2$. 
Next, consider the following states, each of which has an imbalance $Q=4$:
\begin{IEEEeqnarray*}{rCl}
&|1,2,1,2;0,1,0,1\rangle,&
\\[0.2cm]
&|1,2,1,2;0,0,0,2\rangle.&
\end{IEEEeqnarray*}
However, the second state will be energetically more unlikely since it will have a higher $U_s$ repulsion. 
Hence, the most likely states will be those where, for a given $Q$, the atoms are as distributed as possible within each sub-lattice. 
That is, if $Q>0$, then the even sub-lattice should only have $1$'s and $2$'s and the odd sub-lattice should only have $0$'s and $1$'s (and vice-versa for $Q<0$).
We can also reach the same conclusion by imposing that  in the limit $J \to 0$, 
our variational method should reproduce \emph{exactly} the Landau free 
energy~(\ref{landau}). 
That is, we should impose that given the Hamiltonian $H_0$ in Eq.~(\ref{H0}), our states must produce diagonal elements of the form
\begin{equation}\label{EQ}
\epsilon_Q := \langle Q | H_0 | Q \rangle = K f(Q/K) = \frac{U_s |Q|}{2} - \frac{U_\ell Q^2}{K}.
\end{equation}
It is straightforward to show that this will only take place if the above conditions are met. 

The previous analysis still leaves a large ambiguity on which the physically reasonable $Q$ states are. 
In order to gain further insight, let us write the hopping Hamiltonian~(\ref{T}) as 
\begin{equation}\label{T_separation}
\mathcal{T} = - \sqrt{2} J (\mathcal{T}_e + \mathcal{T}_o),
\end{equation}
where 
\begin{equation}\label{Te}
\mathcal{T}_e = \frac{1}{\sqrt{2}} \sum\limits_{i\in e} \sum\limits_{j\in \Gamma_i} b_i^\dagger b_j,
\end{equation}
and $\mathcal{T}_o = \mathcal{T}_e^\dagger$. Here $\Gamma_i$ stands for the neighborhood of site $i$. 
A direct calculation then shows that 
\begin{equation}\label{algebra}
[\Theta,\mathcal{T}_e] = 2 \mathcal{T}_e,\qquad [\Theta,\mathcal{T}_o] = -2 \mathcal{T}_o.
\end{equation}
Thus, $\mathcal{T}_e$ and $\mathcal{T}_o$ function as lowering and raising operators for the imbalance operator. 
The dynamics produced by the particle hopping can thus be interpreted as 
transitions between neighboring $Q$-states; i.e., as a tight-binding in the space of even-odd imbalances.  

This analysis suggests that a reasonable set of $Q$ states is precisely given by those
states that are
generated by the successive application of $\mathcal{T}_e$ and $\mathcal{T}_o$ onto the MI state. 
This process therefore describes the nucleation of CDW 0-2 pairs within the lattice, which has a similar picture to the usual nucleation of domains in thermal phase transitions \cite{Godreche2000,Henkel2010}.
Of course, when we apply these operators, we will also generate states with more than double occupation. 
To amend for this we define
\begin{equation}
\tilde{\mathcal{T}}_e = \mathcal{P}\mathcal{T}_e\mathcal{P},
\end{equation}
where $\mathcal{P}$ is a projection operator onto the subspace of states which 
are at most doubly occupied. 
Note that $\tilde{\mathcal{T}}_e$ no longer satisfies the closed algebra of Eq.~(\ref{algebra}), as is customary of projected operators.
We may then finally define, for $Q>0$,  
\begin{equation}\label{Q_gen}
|Q\rangle  = \frac{1}{\sqrt{A(Q)}} (\tilde{\mathcal{T}}_e)^{Q/2} |MI\rangle,
\end{equation}
where $A(Q)$ is a normalization constant, to be determined.  
For $Q<0$ we replace $\tilde{\mathcal{T}}_e$ with $\tilde{\mathcal{T}}_o$, 
defined in analogous manner.

In order to construct the Sch\"odinger Lagrangian~(\ref{Lagrangian}), we need the matrix elements of the total Hamiltonian~(\ref{H}) in the $Q$ basis. 
The diagonal part $H_0$ is given in Eq.~(\ref{EQ}). 
Next, we must turn to the matrix elements of the hopping term $\mathcal{T}$ in Eq.~(\ref{T_separation}). 
This operator connects $Q$ with $Q\pm 2$. 
In particular, the connection with $Q+2$ will be given by:
\begin{equation}
\langle Q+2 | \mathcal{T}_e | Q \rangle = \langle Q+2 | \tilde{\mathcal{T}}_e 
| Q \rangle = \sqrt{\frac{A(Q+2)}{A(Q)}} \,. 
\end{equation}
Hence, 
\begin{equation}
\langle Q+2 | \mathcal{T} | Q \rangle = - \sqrt{2}J \sqrt{\frac{A(Q+2)}{A(Q)}},
\end{equation}
which is the general formula relating this matrix element with the normalization constant $A(Q)$ of Eq.~(\ref{Q_gen}). 
Analogous formulas can be obtained for $Q<0$, noting that for symmetry 
reasons $A(Q)=A(-Q)$ and $\langle Q+2|\mathcal{T}|Q\rangle = \langle 
-(Q+2)|\mathcal{T}|-Q\rangle$.

\subsection{Distortion of the hopping term}

Unfortunately, computing the normalization constant $A(Q)$ in Eq.~(\ref{Q_gen}) is in general a very difficult task since it depends on the structure of the underlying lattice, that appears in the generator $\tilde{\mathcal{T}}_e$ in Eq.~(\ref{Te}).  
In fact, as shown in Appendix~\ref{app:matching}, this calculation maps exactly 
to the matching  problem, which not only has no analytical solution, but is 
also a $\sharp P$-hard problem to solve numerically (even if we attempt to 
estimate these coefficients by Monte Carlo methods,  the state of the art 
algorithms have a complexity of $\mathcal{O}(K^7\log^4 K)$ 
\cite{Bialecki2010,Jerrum2004}). 
We also mention in passing that this turns out to be the same problem one encounters in computing 
the partition function of the monomer-dimer model or the three dimensional Ising model \cite{Jerrum1987}.

Instead, in order to obtain an analytical treatment, we propose here a deformation of  $\tilde{\mathcal{T}}_e$ to eliminate the underlying lattice structure. 
The deformation is similar to the one routinely 
done in mean-field treatments  and consists in allowing an atom to hop from any  even site to any odd site. 
That is, we deform $\tilde{\mathcal{T}}_e$ to a new hopping operator $\tilde{\mathcal{T}}'_e$ defined as
\begin{equation}\label{distortion}
\tilde{\mathcal{T}}'_e = \frac{1}{\sqrt{2}} \sum\limits_{i\in e} \sum\limits_{j \in o} \mathcal{P}b_i^\dagger b_j  \mathcal{P},
\end{equation}
where the two sums in $e$ and $o$ are now unrestricted. We remark that we make 
this replacement for the definition of the $|Q\rangle$ basis states only, and 
not 
in the Hamiltonian of the system.

With this modified hopping generator, we can now compute the matrix element of the hopping term $\mathcal{T}$. 
To illustrate the procedure, it suffices to consider $Q>0$. 
Then the even sub-lattice will only have occupation 1 or 2 and the odd sub-lattice will only have occupation 1 or 0. 
Let us define $C_e$ as the list of sites in the even sub-lattice which are doubly occupied and $C_o$ as the list of sites in the odd sub-lattice which are empty.
We also define the state $|C_e,C_o\rangle$ as the Fock state with the sites in $C_e$ doubly occupied, the sites in $C_o$ empty and all others singly occupied. 
Then, the set of $Q$ states generated as in Eq.~(\ref{Q_gen}), but with $\tilde{\mathcal{T}}_e'$ instead, may be written as 
\begin{equation}\label{Q_state_final}
|Q\rangle = \frac{1}{\sqrt{A(Q)}} \sum\limits_{\{C_e, C_o\}} |C_e,C_o\rangle.
\end{equation}
The normalization constant $A(Q)$ is now simply the number of states in this list, which is
\begin{equation}
A(Q) = \binom{K/2}{Q/2}^2.
\end{equation}

Now  let us analyze the action of the operator $\mathcal{P}(b_i^\dagger b_j)\mathcal{P}$ in the $Q$-state~(\ref{Q_state_final}), assuming $i \in e$ and $j \in o$. 
This will only contribute if it acts on a pair of sites with $n = 1$. 
Thus, 
\begin{equation}
\mathcal{P}(b_i^\dagger b_j)\mathcal{P}
| C_e, C_o\rangle
= 
\begin{cases}
\sqrt{2} |C_e+i, C_o+j\rangle, 
& i\notin C_e, \quad j \notin C_o \\[0.2cm]
0 & \text{otherwise}
\end{cases},
\end{equation}
where $C_e+i$ is a short-hand notation for the list $C_e$ with the entry $i$ appended to it, and similarly for the other term. 
It then immediately follows that 
\begin{widetext}
\begin{IEEEeqnarray*}{rCl}
\langle Q+2 | \mathcal{P}(b_i^\dagger b_j)\mathcal{P} | Q\rangle &=& 
\frac{1}{\sqrt{A(Q) A(Q+2)}} \sum\limits_{C_e,C_o,C_{e'},C_{o'}} \langle 
C_{e'},C_{o'}| b_i^\dagger b_j |C_e, C_o\rangle		\\[0.2cm]
&=& \frac{\sqrt{2}}{\sqrt{A(Q) A(Q+2)}} \sum\limits_{C_e, C_o} (\text{only 
configurations with } i \notin C_e \text{ and } j \notin C_o).
\end{IEEEeqnarray*}
\end{widetext}
This sum corresponds to the number of ways of distributing $Q/2$ ``2's'' into $K/2-1$ even sites and $Q/2$ ``0's'' into $K/2-1$ odd sites. 
We note that the lists $C_{e/o}$ contain $Q/2$ sites while lists $C'_{e/o}$ 
involve $Q/2 + 1$ sites; we do not make this explicit in the naming of the 
lists to avoid making the notation cumbersome. 
Thus we get
\[
\langle Q+2 | b_i^\dagger b_j | Q\rangle = \sqrt{2} \frac{\binom{K/2-1}{Q/2}^2}{\binom{K/2}{Q/2} \binom{K/2}{Q/2+1}},
\]
which, after simplifying, yields
\begin{equation}
\langle Q+2 | b_i^\dagger b_j | Q\rangle = \sqrt{2} \frac{(K-Q)(Q+2)}{K^2}.
\end{equation}

Note that this result is independent of $(i,j)$, which is a consequence of the distortion introduced in Eq.~(\ref{distortion}), that washes out
all lattice information. 
Consequently, it is now trivial to find the matrix elements of $\mathcal{T}$:
\begin{equation}
\langle Q+2 | \mathcal{T} | Q \rangle =- J \sqrt{2} \frac{(K-Q)(Q+2)}{K^2} 
\sum\limits_{i\in e} \sum\limits_{j\in \Gamma_i} 1 .
\label{eq:tunneling}
\end{equation}
This sum has $\frac{K}{2} \times z$ elements, with $z$ the number of neighbours 
($z=4$ in the two-dimensional case we consider), so we finally obtain
\begin{equation}\label{Ve_mat_elem_pos}
\gamma_Q^+ := \langle Q+2 | \mathcal{T}_e | Q \rangle = -\frac{J z}{\sqrt{2}} 
\, \frac{(K-Q)(Q+2)}{K},
\end{equation}
an equation valid for $Q\geq0$. 
By taking the adjoint of Eq. (\ref{eq:tunneling}) we also immediately get 
\begin{equation}\label{Vo_mat_elem_pos}
\gamma_Q^- := \langle Q-2 | \mathcal{T}_o | Q \rangle =- \frac{J z}{\sqrt{2}}
\frac{(K-Q+2) Q}{K},
\end{equation}
for $Q\geq2$. To obtain the remaining matrix elements we simply note that, 
when the sign of $Q$ changes, the roles of $\mathcal{T}_e$ and 
$\mathcal{T}_o$ are inverted. In summary, we get 
\begin{widetext}
\begin{IEEEeqnarray}{rCl}
\label{gammaP}
\gamma_Q^+ :=\langle Q +2 | H | Q \rangle &=& - \frac{\alpha}{4K } 
\begin{cases}
(K-Q)(Q+2) 	&  Q\geq 0 	\\[0.2cm]
(K- |Q|+2) |Q|  & Q <0 
\end{cases},
\\[0.2cm]
\label{gammaM}
\gamma_Q^- :=\langle Q -2 | H | Q \rangle &=& - \frac{\alpha}{4K } 
\begin{cases}
\displaystyle{(K-Q+2)Q},	&	Q> 0	\\[0.4cm]
\displaystyle{(K-|Q|)(|Q|+2)},	&	Q\leq0
\end{cases},
\end{IEEEeqnarray}
\end{widetext}
where we have defined a rescaled hopping parameter 
\begin{equation}
\alpha = 2 \sqrt{2} z J.
\end{equation} 
The functions $\gamma_Q^\pm$ represent the effective hopping in the space of imbalances and correspond to the off-diagonal entries in the effective Hamiltonian~(\ref{Heff}).

We finish this section by emphasizing that the present choice of $|Q\rangle$-states neglects lattice information, such as dimensionality and number of neighbors. 
One should therefore not expect that this model would reproduce details of the phase diagram depending on these quantities such as, for instance, the location of the MI-SF transition. 
On the other hand, as far as the MI-CDW transition is concerned, this 
dependence should be weak since, in the limit of zero hopping, the lattice 
structure plays no role at all.

%
%
\section{\label{sec:ana}Analysis of the variational Hamiltonian}
%
%

\subsection{Numerical analysis}

\begin{figure}
\centering
\includegraphics[width=0.22\textwidth]{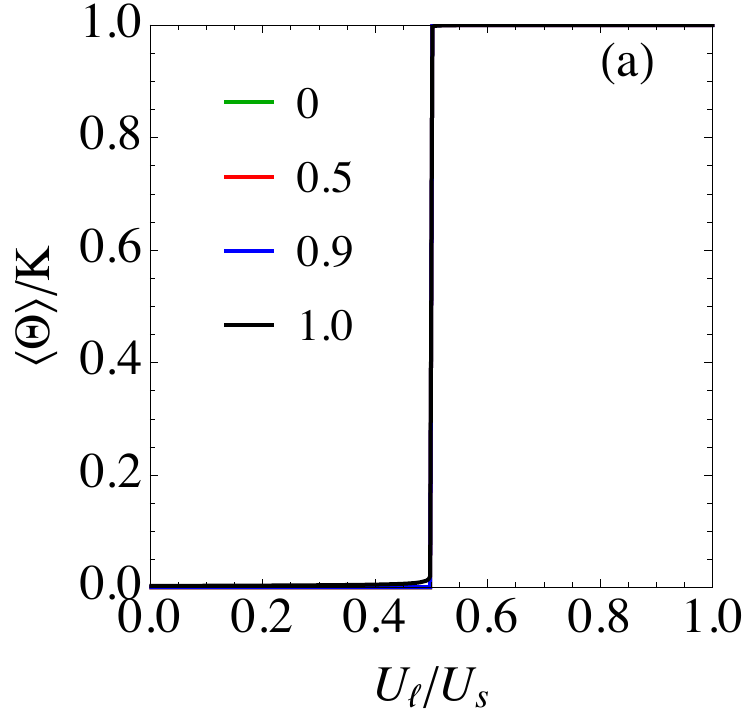}\quad
\includegraphics[width=0.22\textwidth]{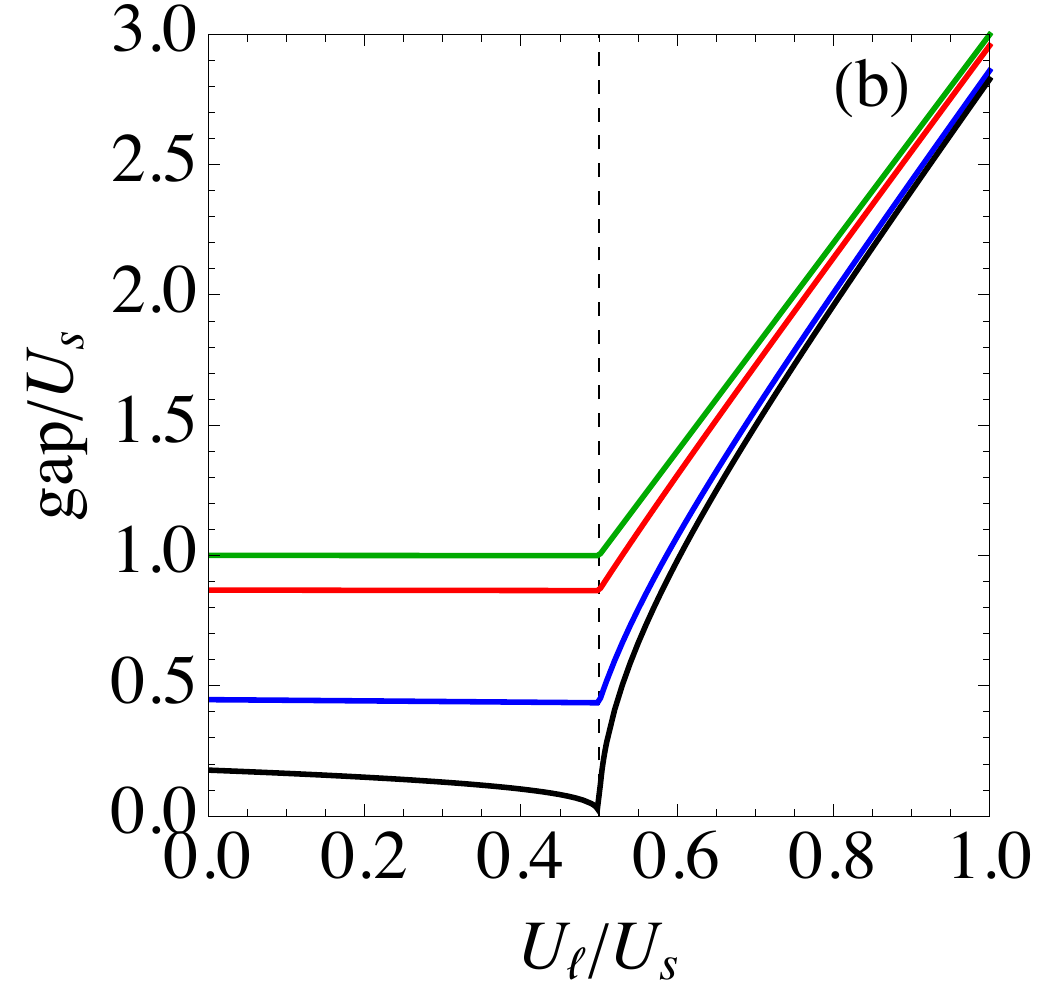}\\
\includegraphics[width=0.22\textwidth]{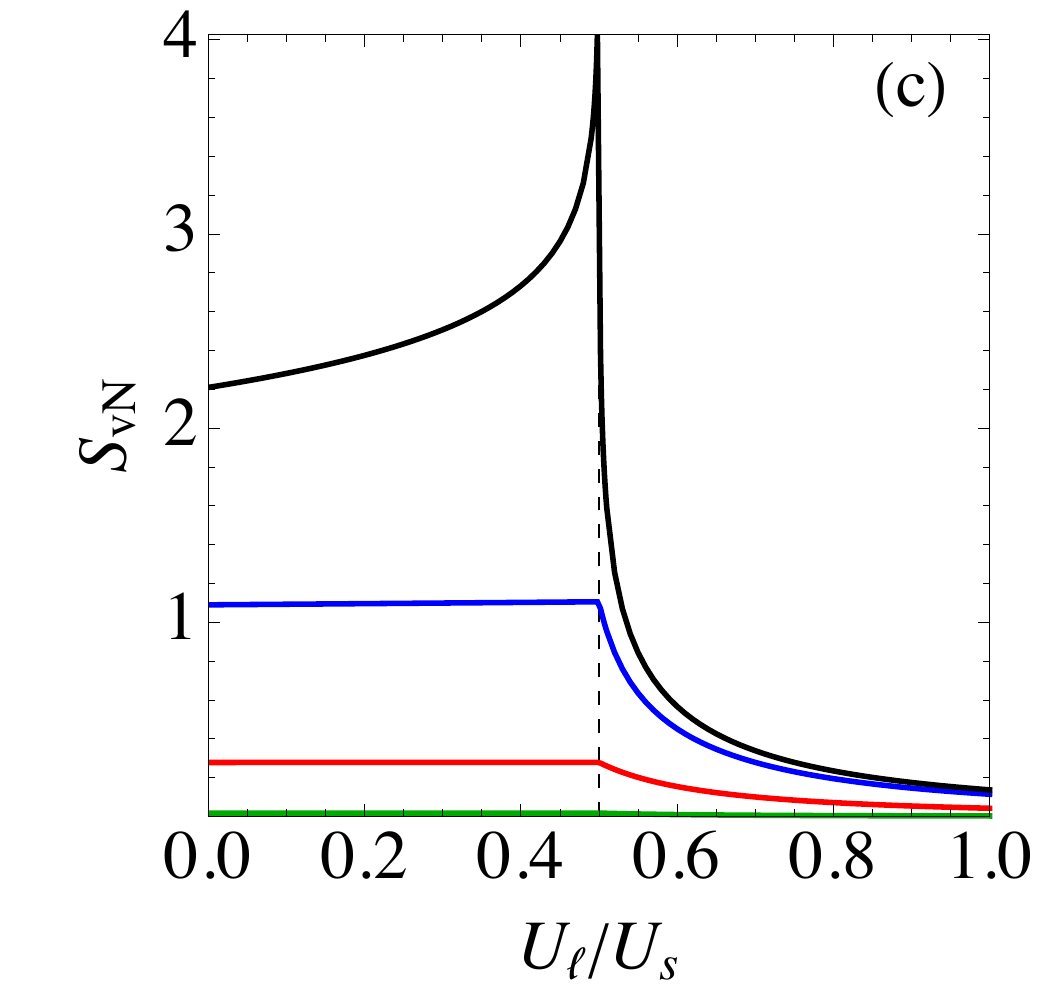}\quad
\includegraphics[width=0.22\textwidth]{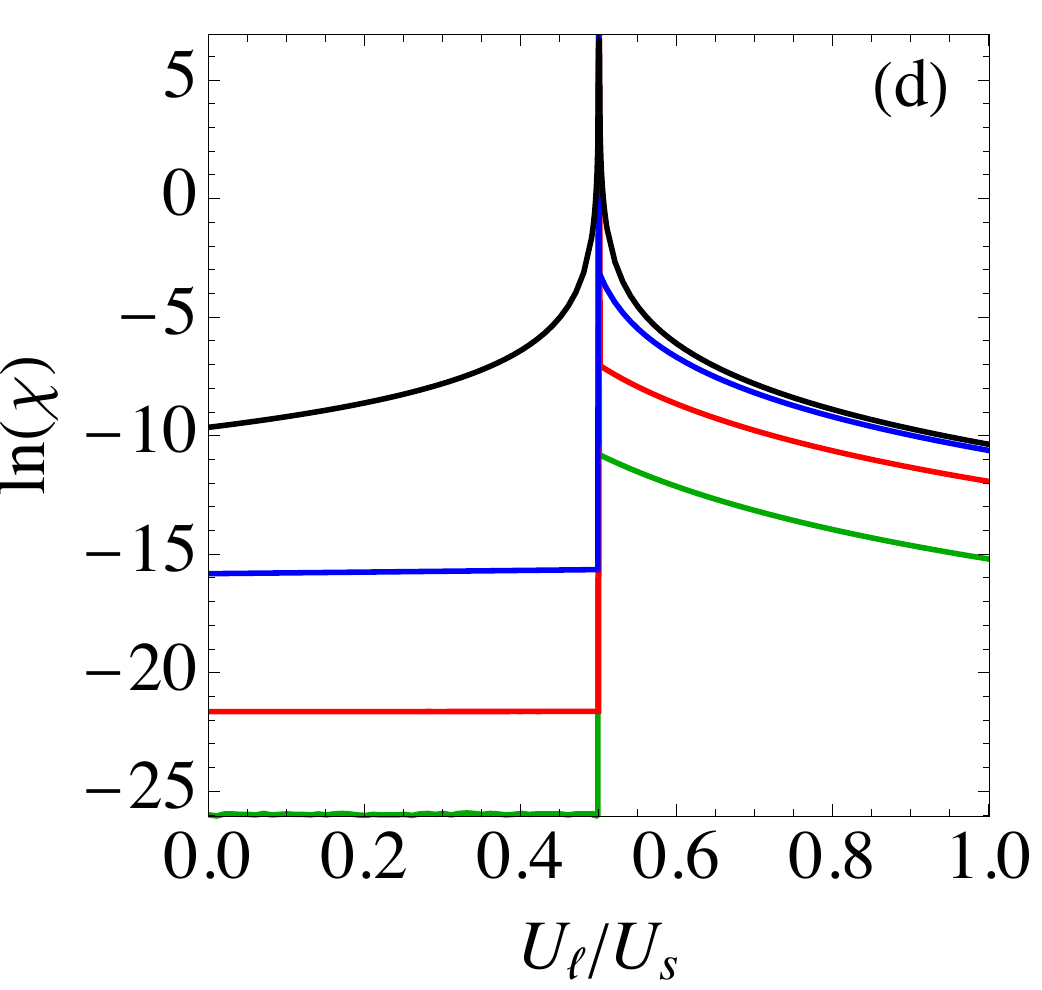}
\caption{\label{fig:props}(Color Online)
Ground-state properties of the variational Hamiltonian, computed as a function of $U_\ell/U_s$ for $K = 2000$ and different values of $\alpha/U_s$, from 0 to 1 [the curve corresponding to $\alpha/U_s = 0$ is the uppermost curve in (b) and the lowermost in (c) and (d)].
The quantities plotted in each figure are: (a) $\langle \Theta \rangle$.
(b) The energy gap between the ground-state and the first excited state. 
(c) The entanglement entropy between the even and the odd sub-lattices. 
(d) The fidelity susceptibility (plotted logarithmically for clarity). 
}
\end{figure}

In this section we study, both numerically and analytically,  the effective 
variational Hamiltonian in Eq.~(\ref{Heff}). 
We begin with a numerical analysis of the ground-state and the excitations, computed by diagonalizing the tridiagonal Hamiltonian~(\ref{Heff}) using standard routines. 
The results are shown in  Fig.~\ref{fig:props} as a function of $U_\ell/U_s$ 
for a fixed $K = 2000$ and different values of the effective hopping term 
$\alpha = 2 \sqrt{2} z J$. We stress that for completeness we include 
results corresponding to all values of $\alpha$. However, our model is only 
built to describe the MI-CDW transition, i.e. the regime of small tunneling 
rates.

In Fig.~\ref{fig:props}(a) we present the average imbalance $\langle \Theta\rangle/K$. 
It shows a discontinuous transition at $U_\ell = U_s/2$, from $0$ to $1$. 
This represents the MI-CDW transition in the limit of small hopping. 
Such a discontinuous  transition continues up until $\alpha = 1$, above which one enters a compressible phase which falls outside the scope 
of this variational model.
A similar behavior is found in the energy gap between the ground-state and the first excited state, illustrated in Fig.~\ref{fig:props}(b).
For $\alpha<1$ the gap is independent of $U_\ell$ in the MI phase and grows linearly in the CDW phase. 
For $\alpha = 1$ the gap closes at $U_\ell = 1$, marking the existence of a
compressible phase (we stress though, that our model is not a good description 
of this regime).

The simplification afforded by our choice of $Q$ states in Eq.~(\ref{Q_state_final}) allows us also to compute the entanglement entropy between the even and the odd sub-lattices, which is defined as
\begin{equation}
S_\text{vN} = - \tr\rho_e \ln \rho_e ,
\end{equation}
where $\rho_e = \tr_o |\psi\rangle\langle \psi|$. 
The Q basis in Eq.~(\ref{Q_state_final}) can  be split as a tensor product of states in the even and odd sub-lattices.
Consequently, it follows from a straightforward calculation that we will have
\begin{equation}
S_\text{vN} = - \sum_Q |\psi_Q|^2 \ln |\psi_Q|^2.
\end{equation}
This is shown in Fig.~\ref{fig:props}(c) for the numerically computed ground-state wave-function.  
The entanglement entropy is taken as a measure of the correlation between the two sub-lattices. 
As expected, we see that the hopping increases their correlation, which is also generally larger in the MI than in the CDW phase. 

Finally,  in Fig.~\ref{fig:props}(d) we compute the fidelity susceptibility  \cite{Wang2015a,Buonsante2007}, 
\begin{equation}
\chi = - \frac{\partial^2}{\partial \delta^2} \ln |\langle \psi(U_\ell) | \psi(U_\ell +\delta)\rangle| \bigg|_{\delta = 0}.
\end{equation}
This quantity is a well-known  indicator of criticality, and indeed for $\alpha 
<1$ it presents a discontinuous jump at the MI-CDW transition, while for 
$\alpha \geq 1$ it presents a peak at the onset of the compressible phases. 
In fact, we find from our simulations that, quite remarkably, the fidelity can 
pinpoint all transitions of the model even for sizes as small as $K = 10$.

From these results we reconstruct the equilibrium phase diagram depicted by the blue (solid) and yellow (dotted) lines in Fig.~\ref{fig:PD}.
Surprisingly, even though our model  was constructed with the intent of describing only the MI-CDW transition, it turns out to capture a similar  phase diagram as that obtained in Ref~\cite{Flottat2017b} using the Gutzwiller approximation. 
Our model does induce, however, some imprecision in the case of the compressible phases. 
First, it cannot distinguish between superfluid and supersolid, which is a consequence of the reduced Hilbert space employed in our variational wave-function. 
Second, it predicts a MI-SF transition occurring at $\alpha/U_s = 1$. 
Since $\alpha = 2 \sqrt{2} z J$, this gives a transition at $U_s/J = 11.31$, 
which is approximately half of the value obtained by mean-field approximations 
or quantum Monte Carlo 
\cite{Fisher1989,Jaksch1998,Sachdev1998,Greiner2002,Flottat2017b,Dogra2016b}.

\subsection{Discrete WKB method}

\begin{figure}
\centering
\includegraphics[width=0.4\textwidth]{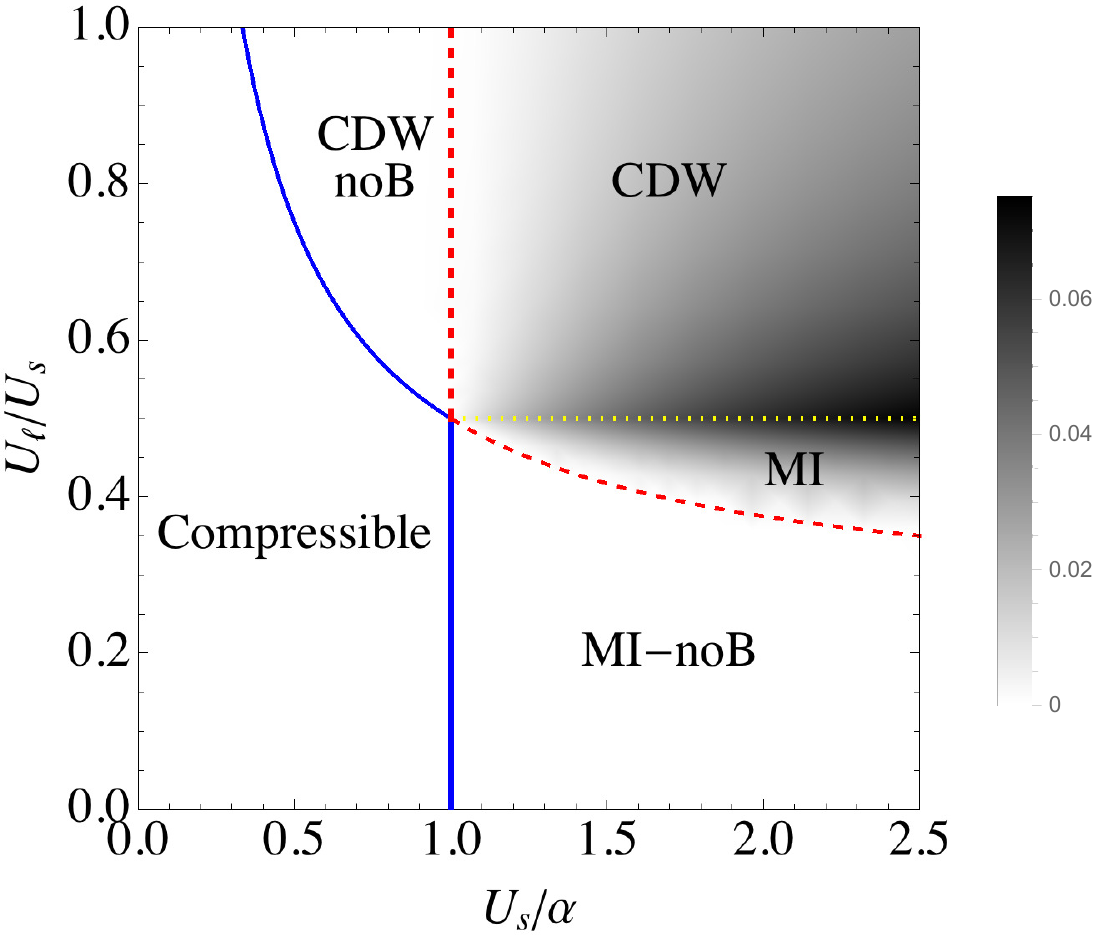}
\caption{\label{fig:PD} 
(Color Online) 
The quantum phase diagram predicted by our variational model. 
The yellow-dotted line represents a first order transition between the MI and CDW phases. 
The blue-solid line represents a second order transition to a compressible phase. 
The red-dashed lines separate the regions where the barrier between the two phases is destroyed. 
The notation ``noB'' means ``no barrier between the two phases''. 
Finally, the gray-scaled density plot depicts the height of the energy barrier (in units of $K U_s$) that the system must overcome to transition between the metastable minimum and the global minimum.
}
\end{figure}

Next we show that the phase diagram in Fig.~\ref{fig:PD} can actually be obtained analytically using a discrete WKB approximation \cite{Braun1993a,Garg1999,Isakov2016}. 
Moreover, this approach will allow us to augment the  phase diagram with additional information on the height of the energy barrier separating the two phases (red lines in Fig.~\ref{fig:PD}). 

\begin{figure}
\centering
\includegraphics[width=0.22\textwidth]{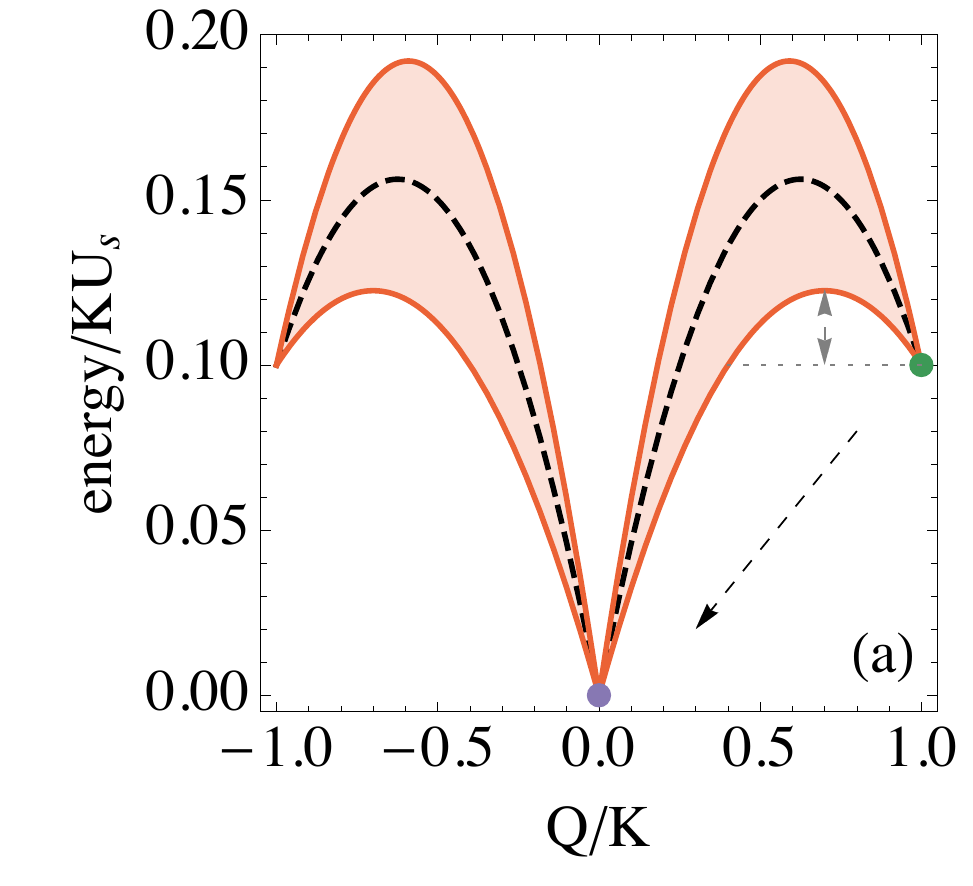}\quad
\includegraphics[width=0.22\textwidth]{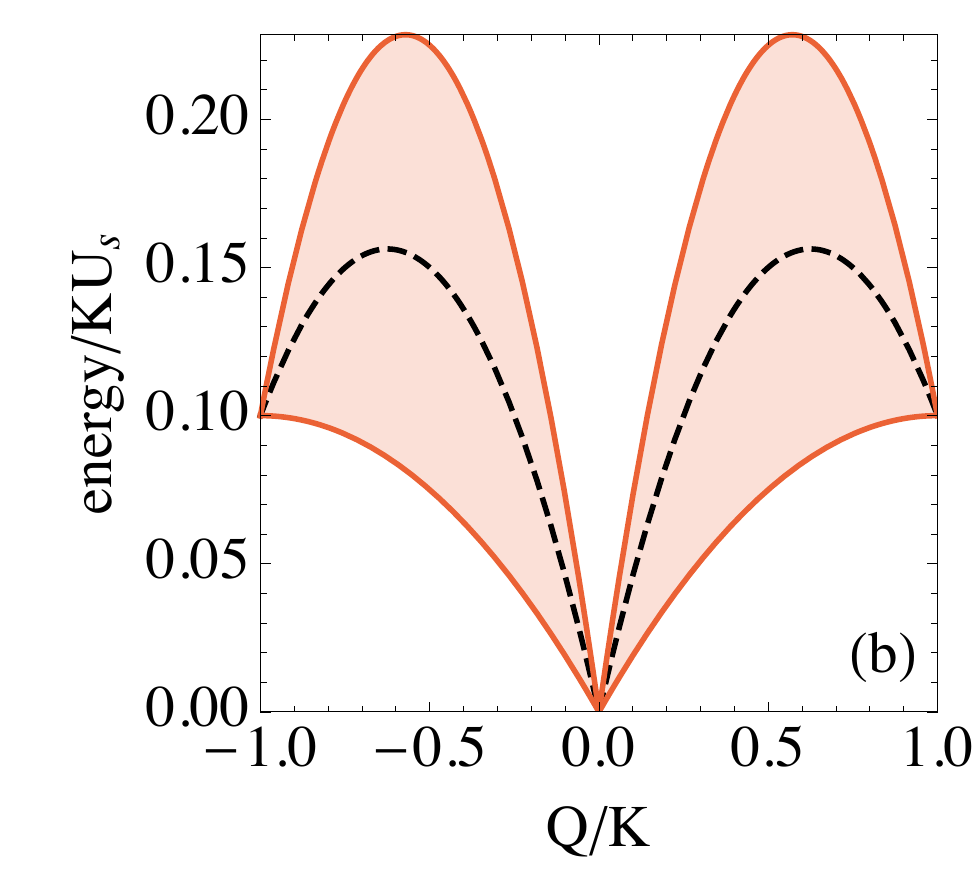}\\
\includegraphics[width=0.22\textwidth]{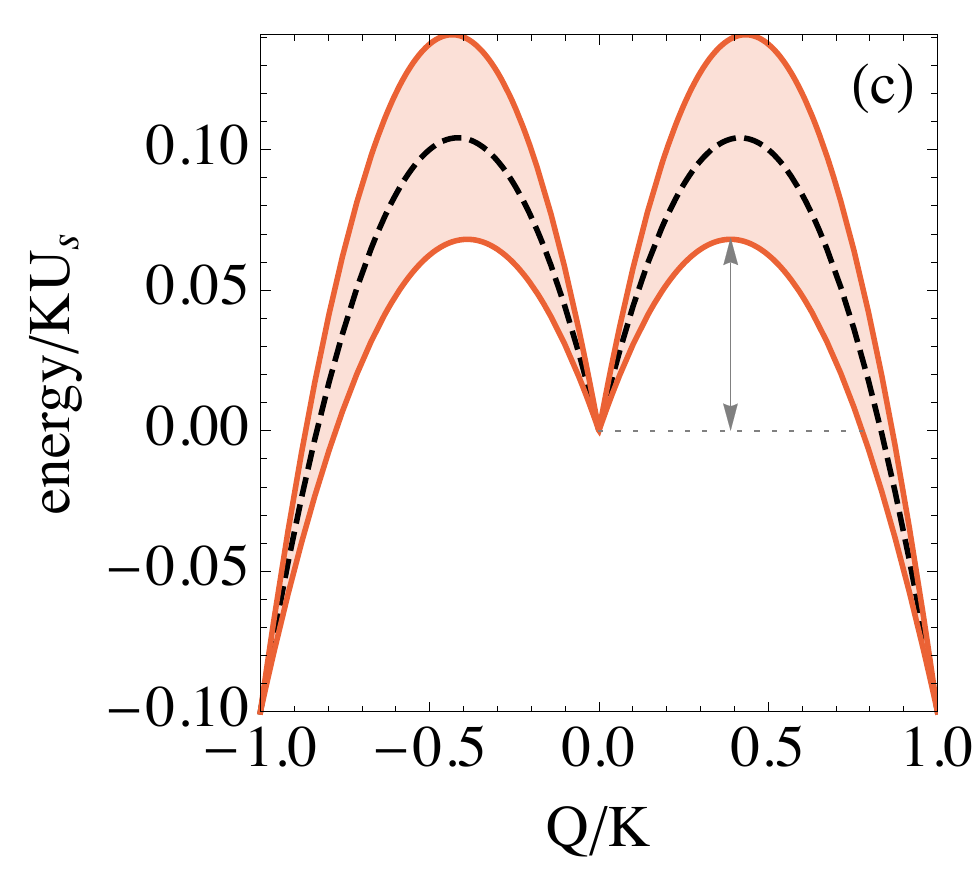}\quad
\includegraphics[width=0.22\textwidth]{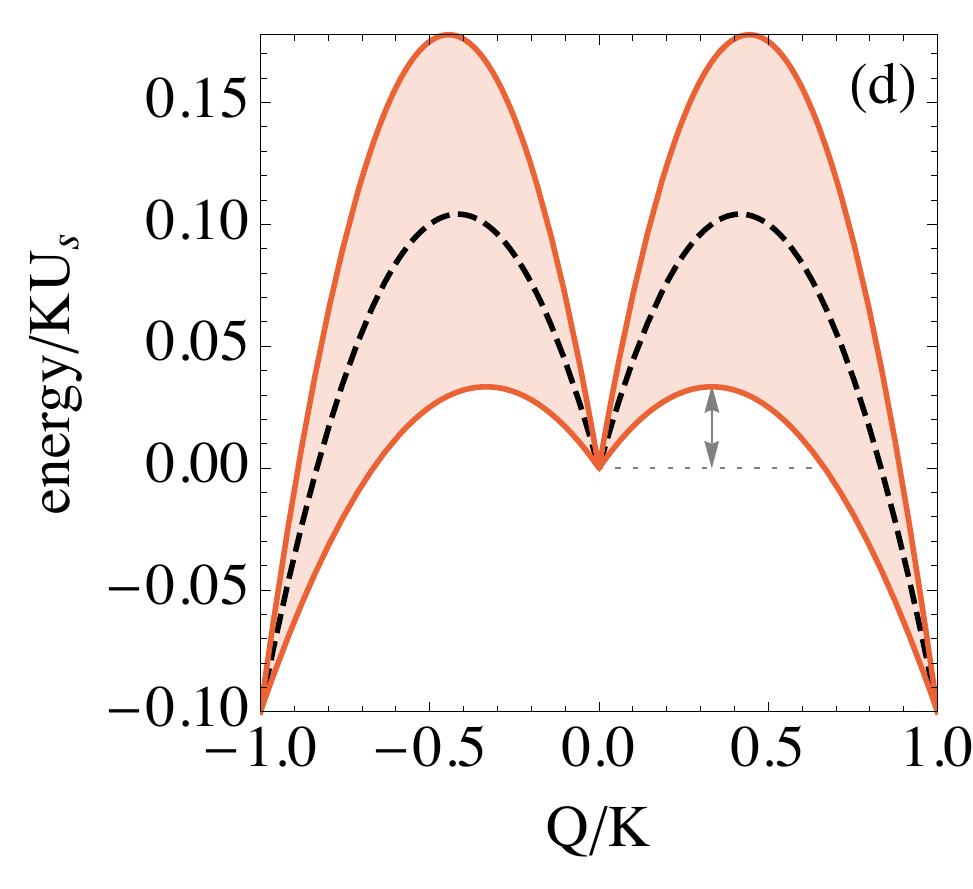}
\caption{\label{fig:WKB_barrier}
(Color Online)
The classically allowed region according to the discrete WKB method. 
The black-dashed line is $\epsilon_Q$, the energy in the absence of hopping (cf. Fig.~\ref{fig:landau}).
The red-solid lines represent the limits $\epsilon_Q \pm 2 \gamma_Q$ [Eq.~(\ref{regions})] and the shaded region between the red lines represent the classically allowed regions.
Finally, the dashed vertical arrows represent the size of the effective barrier the system must overcome when particle hopping is allowed.
(a),(b) $U_\ell/U_s = 0.4$, with $\alpha/U_s = 0.3, 0.6$ respectively.
(c), (d) $U_\ell/U_s = 0.6$ with $\alpha/U_s = 0.3, 0.6$ respectively. 
In (a) we also show the direction of the tunneling from the meta-stable state (green dot) toward the global minimum (purple dot).
The barrier that the system must surmount in this case is no longer given by $\epsilon_Q$, the black-dashed line, but by the lowermost red curve, corresponding to $\epsilon_Q + 2\gamma_Q$.
In (b), on the other hand, the barrier has been completely destroyed by the hopping. 
}
\end{figure}

The essence of the WKB method is to note that the effective Hamiltonian~(\ref{Heff}) closely resembles the tight-binding model describing the hopping of particles through a linear lattice.
The major difference is that, in our case, the hopping terms $\gamma_Q^\pm$ are non-uniform. 
If the hopping was uniform, then it is well known that the eigenstates of the Hamiltonian would be momentum plane waves $e^{i p Q}$ with dispersion relation $E = \epsilon + 2 \gamma \cos(p)$, where $p$ is the momentum and $E$ is the energy.
In our case, even though the hopping is not uniform, in the limit of large $K$ the hopping parameters $\gamma_Q^\pm$ will change slowly with $Q$. 
This motivates us to define $Q$-dependent momenta of the form \cite{Garg1999}:
\begin{equation}\label{pQ}
\cos p(Q) = \frac{E - \epsilon_Q}{2 \gamma_Q},
\end{equation}
where
\begin{equation}
\gamma_Q := \frac{\gamma_Q^+ + \gamma_Q^-}{2} = -\frac{\alpha}{4} \bigg( 1 + |Q| - \frac{Q^2}{K}\bigg),
\end{equation}
which, by construction, is always negative.
The condition that $p(Q)$ should be real determines the classically allowed regions where the eigenvectors of the effective Hamiltonian are oscillatory:
\begin{equation}\label{regions}
\epsilon_Q + 2\gamma_Q \leq E \leq \epsilon_Q - 2\gamma_Q.
\end{equation}
In Fig.~\ref{fig:WKB_barrier} we present these classically allowed regions for different combinations of $U_\ell$ and $\alpha$. 

With the classically allowed regions properly identified, we may now finally make the link with the meta-stability in Fig.~\ref{fig:hysteresis} and understand exactly how the particle hopping facilitates the transition between metastable states, which is the primary goal of this paper.
In Fig.~\ref{fig:WKB_barrier}(a), for instance, we present results for $U_\ell/U_s = 0.4$ and $\alpha/U_s = 0.3$. 
In this case the global minimum is the MI phase, whereas the CDW configurations represent metastable states. 
In the absence of hopping ($\alpha = 0$), the energy barrier that must be surmounted is represented by the black, dashed curve in Fig.~\ref{fig:WKB_barrier}(a). 
When hopping is present, however, the entire shaded region within the two red curves [cf. Eq.~(\ref{regions})] becomes classically allowed, meaning that the barrier that the system will actually have to surmount will be that given by the lowest of the two red curves, which is substantially smaller than the original barrier. 
In fact, in Fig.~\ref{fig:WKB_barrier}(b) we present curves for the same value of $U_\ell/U_s$ but a larger value of $\alpha$. 
In this case, as can be seen, the hopping has completely destroyed the barrier between the two minima.

We now use this information to construct the phase diagram in Fig.~\ref{fig:PD} 
and augment it with information on the barrier height. 
From  Eq.~(\ref{regions}), we see that the quantity determining the energy barrier will be the lower bound
\begin{IEEEeqnarray}{rCl}
\mathcal{E}(Q) &=& \epsilon_Q + 2\gamma_Q \\[0.2cm]
&=& - \frac{\alpha}{2} + \frac{|Q|}{2} (U_s-\alpha) - 
\frac{Q^2}{2K}(2U_l-\alpha),
\label{mathcal_barrier}
\end{IEEEeqnarray}
which correspond to the bottom red curves in Fig.~\ref{fig:WKB_barrier}. 
Our first goal is to use this to establish in which cases a barrier exist.
For instance, in the examples of Fig.~\ref{fig:WKB_barrier}, it exist only in figures (a), (c) and (d), where they are signaled by vertical arrows. 
First, a necessary (but not sufficient) condition can be easily found by noting that 
Eq.~(\ref{mathcal_barrier}) is a parabola in $Q$, so that for a barrier to exist the coefficient of the quadratic term must be negative, hence implying $\alpha<2U_\ell$. Above this value we are always in a compressible phase and thus no barrier exists.

Next, focusing on $Q>0$, we note that the extremum of $\mathcal{E}$ occurs at
\begin{equation}
\frac{Q^*}{K} = \frac{1}{2} (U_s-\alpha)/(2 U_\ell - \alpha).
\end{equation}
We must then demand that this extremum exists in the interval 
$Q^*/K \in [0,1]$; the limiting cases turn out to be: 
\begin{equation}
\alpha_1 = U_s\quad \text{ and }\quad \alpha_2 = 4U_\ell - U_s.
\end{equation}
The meaning of these equations change depending on the value of $U_\ell$ since 
the point $Q^*/K$ will only be a maximum ($\mathcal{E}''<0$) provided $\alpha < 
2U_\ell$.
\begin{itemize}

\item If $U_\ell > U_s/2$ (CDW phase) then the line $\alpha_1 = U_s$ determines the region where the barrier seizes to exist, which is denoted by the red (dashed) vertical line in Fig.~\ref{fig:PD}.
Conversely, the line $\alpha_2 = 4U_\ell - U_s$ denotes the line after which the extremum becomes a minimum, which signals the transition to a compressible phase. This is denoted by the curved blue line in Fig.~\ref{fig:PD}, which is defined for $U_\ell > U_s/2$. 

\item If $U_\ell < U_s/2$ (MI phase) then the roles are inverted. The line $\alpha_1 = U_s$ determines the transition to a compressible phase (blue line in Fig.~\ref{fig:PD}) and the line $\alpha_2 = 4U_\ell - U_s$ denotes the region below which the barrier vanishes (red line in Fig.~\ref{fig:PD}).
 
\end{itemize}
 
We may  also construct the energy barrier height that must be overcome in order to transition from a metastable minimum to the true global minimum. 
If we are in the CDW phase then the barrier height will be given by 
\begin{equation}
\Delta \mathcal{E}_\text{CDW} = \mathcal{E}(Q^*) - \mathcal{E}(0)
=\frac{K}{8} \frac{(U_s-\alpha)^2}{2U_\ell -\alpha}.
\end{equation} 
Conversely, if we are in the MI phase, we will have
\begin{equation}
\Delta \mathcal{E}_\text{MI} = \mathcal{E}(Q^*) - \mathcal{E}(K)
=\frac{K}{8} \frac{(\alpha+ U_s - 4 U_\ell)^2}{2U_\ell -\alpha}.
\end{equation} 
These results are shown in the form of density plots in Fig.~\ref{fig:PD}. 
Thus, we see that from the discrete WKB method we may not only compute the entire equilibrium phase diagram, but also determine the regions where barriers between minima exist and the height of these barriers.

Finally, we use the discrete WKB method to address the problem of macroscopic quantum tunneling from the metastable state toward the true global minimum. 
For instance, in the example of Fig.~\ref{fig:WKB_barrier}(a), we shall consider the tunneling  from the CDW state $Q= K$ toward the MI state $Q=0$, and vice-versa in the case of Figs.~\ref{fig:WKB_barrier}(c) and (d).
Following \cite{Braun1993a,Garg1999,Isakov2016} we write the tunneling probability as 
\begin{equation}
\mathcal{T} = e^{- 2 K  \mathcal{A}},
\end{equation}
where 
\begin{equation}\label{A}
\mathcal{A} =  \frac{1}{K} \int\limits_{Q_1}^{Q_2} |p(Q)| \ud Q,
\end{equation}
with $p(Q)$  given in Eq.~(\ref{pQ}) and $Q_{1,2}$ being the classical turning points of the dynamics. 
They are indicated in Figs.~\ref{fig:WKB_barrier}(a), (c) and (d) by horizontal dashed lines.
From $\mathcal{T}$ one may estimate the average life-time as $\tau \sim 1/\mathcal{T}$, which can unfortunately only be done up to a pre-factor that is not easily computed. 

The integral in~(\ref{A}) can be computed analytically, even though the result is somewhat cumbersome.
To do so, we treat separately the cases $U_\ell < U_s/2$ and $U_\ell > U_s/2$, since the turning points $Q_{1,2}$ are different in each case. 
Let us start with $U_\ell < U_s/2$, for which the global minimum is the MI phase.
This situation is illustrated in Fig.~\ref{fig:WKB_barrier}(a).
The classical turning points $Q_{1}$ and $Q_2$ are precisely the end-points of the grey dashed line. 
Thus  $Q_2 = K$,  whereas $Q_1$ is the solution of $\mathcal{E}(Q_1) = \mathcal{E}(Q_2)$, which reads  $Q_1 = K(U_s-2U_\ell)/(2U_\ell -\alpha)$.
The integral then yields
\begin{widetext}
\begin{equation}\label{A2}
\mathcal{A} = 
\sech^{-1}\bigg(\frac{\alpha}{4U_\ell - U_s}\bigg) +
\frac{U_s - 2 U_\ell}{\sqrt{4 U_\ell^2 - \alpha^2}}\ln\bigg\{
\frac{\alpha(U_s - 2 U_\ell)}{2 U_\ell U_s - 8 U_\ell^2 + \alpha^2- \sqrt{[(4U_\ell - U_s )^2- \alpha^2](4U_\ell^2 - \alpha^2)}}
\bigg\}.
\end{equation}
We proceed similarly for $U_\ell > U_s/2$. 
In this case the turning points are $Q_1 = 0$ and $Q_2 = K (U_s-\alpha)/(2U_\ell - \alpha)$, which yields 
\begin{equation}\label{A1}
\mathcal{A} = 
\ln\bigg[\frac{U_s + \sqrt{U_s^2 - \alpha^2}}{U_s}\bigg]
+ \frac{U_s - 2 U_\ell}{\sqrt{4 U_\ell^2 - \alpha^2}}
\ln\bigg\{\frac{\alpha(U_s - 2 U_\ell)}{\alpha^2 -2 U_s U_\ell + \sqrt{(4 U_\ell^2-\alpha^2)(U_s^2-\alpha^2 )}}\bigg\}.
\end{equation}
\end{widetext}
These results  simplify considerably in the limit $U_\ell = U_s/2$, which is when all energy minima have the same energy. 
In this case they both tend to
\begin{equation}
\mathcal{A} =\cosh^{-1}\bigg(\frac{U_s}{\alpha}\bigg).
\end{equation}
Thus, we see that when $\alpha\to 0$ the amplitude $\mathcal{A}$ diverges and, consequently, the tunneling probability tends to zero, as expected. 
Conversely, when $\alpha \geq U_s$ the tunneling amplitude is identically zero, as there is no energy barrier to surmount. 
The same is also true, for instance, in the example of Fig.~\ref{fig:WKB_barrier}(b), although that is not so readily seen from Eq.~(\ref{A2}).

We can compare these results directly with the life-time of thermally assisted transitions. 
From the standard Arrhenius theory, the average life-time for thermal transitions should be proportional to $e^{\Delta E/k_B T}$, where $\Delta E$ is the height of the energy barrier. 
In the case $U_\ell = U_s/2$, the height of the barrier is $\Delta E = K (U_s - \alpha)/8$. 
Thus, by analyzing how the average life-time of the metastable states scale with the hopping parameter $\alpha$, it should be possible to infer the relative contribution of quantum tunneling versus thermally assisted transitions. 
The method used here only allows us to estimate the main exponential dependence, thus making it difficult to directly compare with experiment. 
A more thorough analysis of the tunneling time will be the subject of a future publication.

%
%
\section{\label{sec:con}Conclusions}
%
%

In summary, we have discussed the role of quantum fluctuations, represented by the local tunneling of atoms in an optical lattice, in the phase reconfiguration of a first order quantum phase transition. 
Within the energy landscape picture provided by the discrete WKB method, we find that the tunneling acts to lower the energy barrier separating the two phases, hence facilitating the transition between a metastable minimum and the true global minimum. 
Our approach relied on a choice of variational states describing the relative imbalances between the even and odd sub-lattices. 
As a result, we found that the tunneling acts as a local tight-binding in the space of imbalances, causing transitions between neighboring states. 
One may thus interpret the tunneling of atoms as the generators of CDW 0-2 pairs within the lattice. 
These pairs should then form domains which eventually nucleate into a new phase. 
It is our hope that our analysis sheds further light on the non-equilibrium mechanisms responsible for phase reconfigurations in quantum critical systems.

\begin{acknowledgements}

The authors acknowledge fruitful discussions with G. Morigi, L. Hruby, N. Dogra, M. Landini, T. Donner and T. Esslinger.
GTL acknowledges the financial support from grant number 2016/08721-7 from the S\~ao Paulo Research Foundation (FAPESP).
GTL also acknowledges fruitful discussions with E. Andrade, M. J. de Oliveira, S. Salinas and A. P. Vieira. 
SW acknowledges useful discussions with M. Henkel. SW is grateful to the ‘Statistical Physics Group’ at University
of S\~ao Paulo, Brazil and the Group ‘Rechnergest\"utzte Physik der Werkstoffe’ at ETH Z\"urich,
Switzerland, for their warm hospitality and to UFA-DFH for financial support through grant
CT-42-14-II.

\end{acknowledgements}

\appendix

\section{\label{app:matching}Mapping to the matching problem}

In this appendix we show that the task of computing the normalization constants $A(Q)$ in Eq.~(\ref{Q_gen}) can be mapped into a matching problem widely studied in computational combinatorics \cite{Jerrum1987}. 
Let $c_i$ be the operators
\begin{equation}\label{c-def}
c_i
= 
\begin{cases}
b_i, 
& i\in o\\[0.2cm]
\frac{1}{\sqrt{2}}b_i^\dagger, & i \in e
\end{cases},
\end{equation}
and, for any set $S$ of lattice sites, define an operator  $\mathcal{D}_S$ as
\begin{equation}\label{D-def}
\mathcal{D}_S = \mathcal{P}\prod_{i\in S} c_i\mathcal{P}.
\end{equation}
States created by the action of $\mathcal{D}_S$ onto the MI state are orthonormal. 
That is, if $|S\rangle \equiv \mathcal{D}_S|MI\rangle$ then $\langle S | S' \rangle = \delta_{S,S'}$
Moreover, the $\mathcal{D}$ operators obey the algebra
\begin{IEEEeqnarray*}{rCl}
[\mathcal{D}_X, \mathcal{D}_Y] &=& 0,
\\[0.2cm]
\mathcal{D}_X \mathcal{D}_Y &=& 0, \mbox{if } X\cap Y \neq \varnothing ,
\\[0.2cm]
\mathcal{D}_X \mathcal{D}_Y &=& \mathcal{D}_{X\cup Y}, \mbox{if } X\cap Y = \varnothing .
\end{IEEEeqnarray*}

\begin{figure}
\centering
\vspace{0.2cm}
\includegraphics[width=0.5\textwidth]{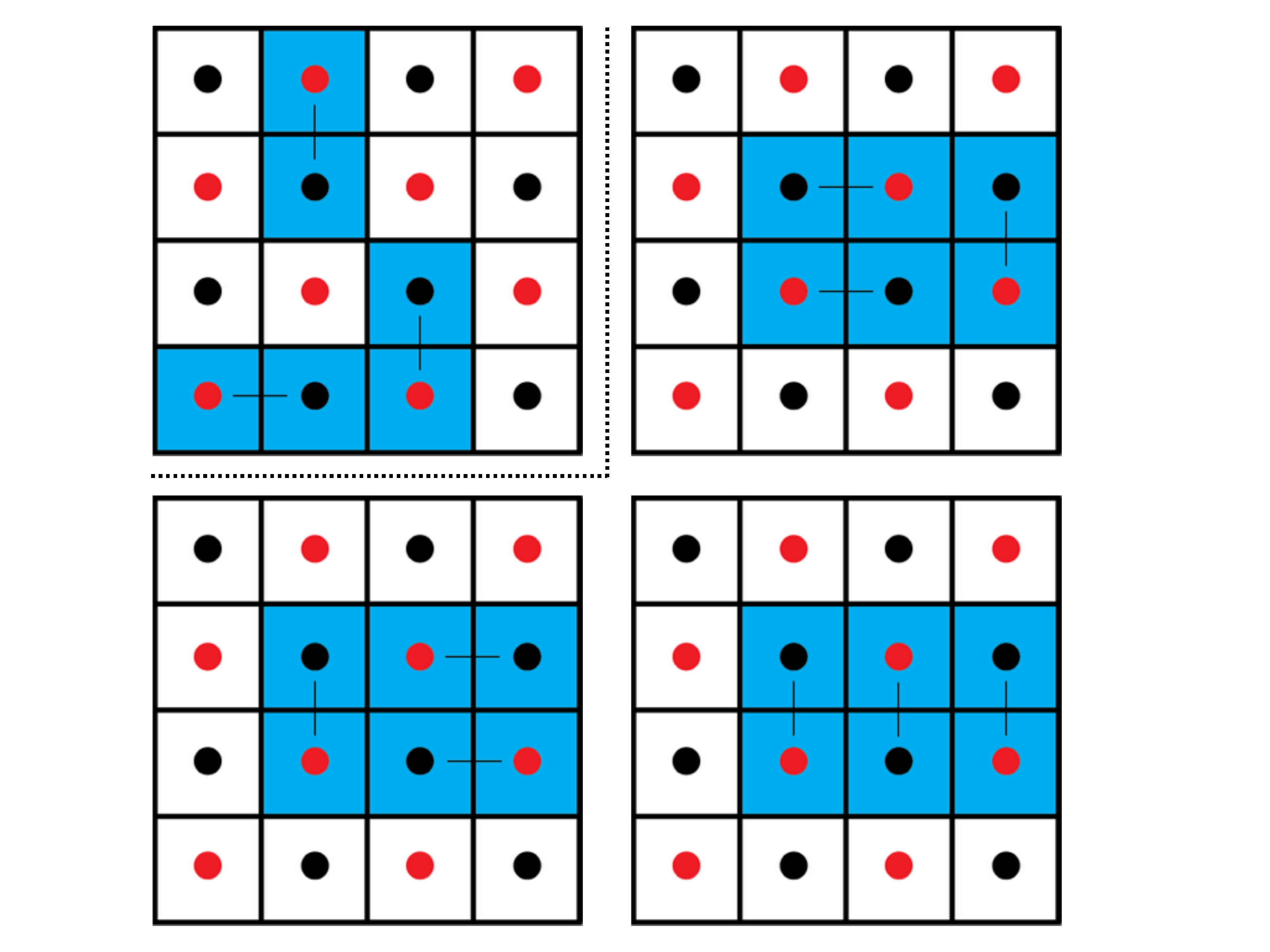}
\caption{\label{fig:ex-C-mu} 
(Color Online)
A $4\times 4$ lattice with two examples of different elements of $\mathcal{C}_3$, the sets of $2\times 3 =6$ sites of the lattice which can be formed from 3 nearest neighbor pairs with no overlap. 
Black and red (light grey) dots represent sites on the even and odd sub-lattices and the shaded blue regions represent the sets in question.
Connecting lines represent the different pairings between even-odd pairs.
On the top left we have an element that can be covered by a triple of nearest neighbor pairs in only one way (hence $\mu = 1$). 
The three other panels show elements of $\mathcal{C}_3$ which can be covered in 3 different ways  so that $\mu=3$. 
 Note that since only neighboring pairs with no overlaps are allowed in elements $\mathcal{R}$ of $\mathcal{C}_n$ it follows that all $\mathcal{R}$ cover an equal number of even and odd sites.
}
\end{figure}

In terms of these new operations, we may write the hopping generator  $\tilde{\mathcal{T}}_e$ as
\begin{equation}\label{Te-D}
\tilde{\mathcal{T}}_e = \sum\limits_{\langle i,j \rangle} \mathcal{D}_{i,j}.
\end{equation}
Hence
\begin{equation}\label{Te-n}
\tilde{\mathcal{T}}_e^n = \sum\limits_{\mathcal{R} \in \mathcal{C}_n} (n!)\mu_{\mathcal{R}}\mathcal{D}_{\mathcal{R}},
\end{equation}
where $\mathcal{C}_n$ is the set containing all sets of $2n$ sites in the lattice, that can be formed from $n$ nearest neighbor pairs with no overlaps and $\mu_{\mathcal{R}}$ is the number of unordered ways to cover $\mathcal{R}$ with such pairs (known as the number of perfect matchings of the set $\mathcal{R}$). We illustrate $\mathcal{C}_n$ and $\mu_{\mathcal{R}}$ with some examples in Fig.~\ref{fig:ex-C-mu}.


From Eqs.~(\ref{Q_gen}) and (\ref{Te-n}) we have
\begin{equation}
|Q\rangle = \frac{1}{\sqrt{A(Q)}} \sum\limits_{\mathcal{R} \in \mathcal{C}_{\nicefrac{Q}{2}}} \left(\nicefrac{Q}{2}\right)!\mu_{\mathcal{R}}|\mathcal{R}\rangle,
\end{equation}
from where we can derive $A(Q)$:
\begin{equation}\label{Aq-matches}
A(Q) = \left(\nicefrac{Q}{2}\right)!^2 \sum\limits_{\mathcal{R} \in \mathcal{C}_{\nicefrac{Q}{2}}} \mu_{\mathcal{R}}^2.
\end{equation}
Calculating the exact expression (\ref{Aq-matches}) for all the needed values of $Q$  involves being able to calculate the total number of matchings in a square lattice:
\[
\sum_n\sum\limits_{\mathcal{R} \in \mathcal{C}_n} \mu_{\mathcal{R}},
\]
which is a $\sharp P$-hard problem, making it unfeasible even numerically.
Even if we relax our requirements and attempt to only estimate these quantities by Monte Carlo methods, the state of the art algorithm is still $\mathcal{O}(K^7\log^4 K)$ \cite{Bialecki2010,Jerrum2004}.

\bibliography{/Users/gtlandi/Documents/library}
\end{document}